# Unusual ferrimagnetism in CaFe$_2$O$_4$

Hiroki Ueda[1,†,‡,*], Elizabeth Skoropata[1,†,*], Cinthia Piamonteze[1], Nazaret Ortiz Hernández[1], Max Burian[1], Yoshikazu Tanaka[2], Christine Klauser[3], Silvia Damerio[4,#], Beatriz Noheda[4,5], and Urs Staub[1,*]

[1] *Swiss Light Source, Paul Scherrer Institute, 5232 Villigen-PSI, Switzerland.*

[2] *RIKEN SPring-8 Center, Sayo, Hyogo 679-5148, Japan.*

[3] *Laboratory for Neutron and Muon Instrumentation, Paul Scherrer Institute, 5232 Villigen-PSI, Switzerland.*

[4] *Zernike Institute for Advanced Materials, University of Groningen, 9747AG- Groningen, Netherlands.*

[5] *CogniGron Center, University of Groningen, 9747AG- Groningen, Netherlds.*

**Abstract**: Incomplete cancellation of collinear antiparallel spins gives rise to ferrimagnetism. Even if the oppositely polarized spins are owing to the equal number of a single magnetic element having the same valence state, in principle, a ferrimagnetic state can still arise from the crystallographic inequivalence of the host ions. However, experimental identification of such a state as "ferrimagnetic" is not straightforward because of the tiny magnitude expected for *M* and the requirement for a sophisticated technique to differentiate similar magnetic sites. We report a synchrotron-based resonant x-ray investigation at the Fe $L_{2,3}$ edges on an epitaxial film of CaFe$_2$O$_4$, which exhibits two magnetic phases with similar energies. We find that while one phase of CaFe$_2$O$_4$ is antiferromagnetic, the other one is "ferrimagnetic" with an antiparallel arrangement of an equal number of spins between two distinct crystallographic sites with very similar local coordination environments. Our results further indicate two distinct origins of an overall minute *M*; one is intrinsic, from distinct Fe$^{3+}$ sites, and the other one is extrinsic, arising from defective Fe$^{2+}$ likely forming weakly-coupled ferrimagnetic clusters. These two origins are uncorrelated and have very different coercive fields. Hence, this work provides a direct experimental demonstration of "ferrimagnetism" solely due to crystallographic inequivalence of the Fe$^{3+}$ as the origin of the weak *M* of CaFe$_2$O$_4$.

† These authors contributed equally to this work.

‡ Present address: SwissFEL, Paul Scherrer Institute, 5232 Villigen-PSI, Switzerland.

# Present address: Institut de Ciència de Materials de Barcelona, ICMAB-CSIC, Campus UAB, 08193 Bellaterra, Spain.

* Correspondence authors: hiroki.ueda@psi.ch, elizabeth.skoropata@psi.ch, and urs.staub@psi.ch

Ferrimagnetism is a type of magnetic order in which the populations or amplitudes of oppositely-polarized spins are different, resulting in net magnetization ($M$) [1]. Antiparallel spins from different constituent magnetic elements [2], different valences of a magnetic element [1], and/or different crystallographic sites unbalancing the numbers of antiparallel spins [3] cause a ferrimagnetic order with a relatively large $M$. However, what if the number of parallel spins and antiparallel spins, both from a single magnetic element with the same valence, is the same but the magnetic sites that host them are crystallographically inequivalent? From the symmetry point of view, such a state breaks the time-reversal symmetry as ferrimagnets. Similar but distinct local coordination environments can result in different sizes of magnetic moments, even if the host ions have the same valence. However, experimental demonstration of such a state as being ferrimagnetic is not straightforward, as it requires independently quantifying two almost equivalent sites and grasping evidence that the sublattices host $M$, which is comparably as small as a potential impurity contribution. Tiny $M$ displays a weak response to applied magnetic fields ($H$) and thus, naturally results in a large coercive field. Hence, such a "ferrimagnetic" state possesses two reversible states but can be robust towards $H$.

CaFe$_2$O$_4$ is a unique material exhibiting two distinct magnetic phases with a collinear spin arrangement, both recognized as antiferromagnetic (AFM) [4,5]. The overall magnetism comes from two Fe sites, Fe1 and Fe2, having octahedral coordination with six surrounding $O^{2-}$. Both sites possess the formal oxidation state of 3+ ($S = 5/2$) and are located on Wyckoff positions type $4c$ of the orthorhombic space group $Pnma$ (#62), which does not have a space-inversion center. Namely, the two Fe sites have nominally the same valence, and there are exactly the same number of atoms of each type in a unit cell. Fe1 has a more distorted octahedral coordination than Fe2 [6]. The sites form a zigzag chain along [001] in a sequence of Fe1-Fe1-Fe2-Fe2 as seen in Fig. 1(a) or 1(b), and thus, two intra-chain magnetic interactions exist: between two different Fe sites and between two Fe of the same type. The super-exchange interactions among $Fe^{3+}$ depend on the bond angles of Fe-O-Fe. The magnetic interactions between the same sites of $Fe^{3+}$ along the [001] chain are weak because the bond angles are close to 90°. The weak intra-chain interactions result in two energetically almost degenerate spin configurations that have either ferromagnetic (FM) or AFM patterns along [001] for $Fe^{3+}$ located at the same site [see Figs. 1(a) and 1(b), respectively]. The former phase is called $A$ phase [magnetic space group: $Pn'ma'$ (#62.448)] while the latter is called $B$ phase [magnetic space group: $Pnma'$ (#62.445)] with spins pointing along [010] but a different stacking sequence along [001]: up-up-down-down and up-down-up-down, respectively. The $B$ phase is the magnetic ground state since the exchange interactions are AFM, while the $A$ phase is metastable due to single-ion anisotropy emerging by mixing higher-energy multiplet states [7]. The local FM spin configuration in the $A$ phase (up-up) can be found as the antiphase domain boundary of the AFM $B$ phase [up-down-up (domain 1) and up-down-up (domain2)].

CaFe$_2$O$_4$ shows a phase transition to the $B$ phase at $T_N{}^B \approx 200$ K upon cooling and coexistence of the $A$ and $B$ phases on further cooling below $T_N{}^A \approx 175$ K [8]. The temperature range of the phase coexistence persists down to the lowest temperatures for single-crystal samples [8] while remains down to $T^* \approx 130$ K for powder samples, below which only the $A$ phase is present [9], as visualized in Fig. 1(c). It is proposed that the $A$ phase is more stable in powder samples than in single-crystals samples because of the strain introduced by grinding [7]. Interestingly, despite the believed AFM nature, a small remanent $M$ of ~0.02 $\mu_B$, which points along [010], summed over eight $Fe^{3+}$ in the unit cell [10] is consistently reported in $CaFe_2O_4$, which onsets at around 185 K. Despite more than a half-century of investigation [4,5], the origin of the small $M$ remains unclear. Previous studies based on magnetometry and neutron scattering proposed contributions to $M$ (i) from $Fe^{2+}$-induced ferromagnetic clusters due to oxygen vacancies [11] and (ii) from domain boundaries of an AFM magnetic structure that can be locally FM [12]. Magnetic diffuse scattering experiments have attributed $M$ to the presence of FM domain boundaries which also display an enhancement in $H$ resulting from an increase in domain density [8]. Recently, uncoupled spins from the AFM domains have been discussed as a possible origin, based on spin Hall magnetoresistance measurements [13]. Note that the small $M$ is not due to a spin canting since $M$ and the Néel vector are both along [010] [12].

As found in Fig. 1(a), all Fe1 (Fe2) spins in the $A$ phase point in the same direction but are antiparallel between the two sites. Hence, whereas the $B$ phase is AFM, the $A$ phase with the up-up-down-down configuration matching the periodicity of the lattice displays a "ferrimagnetic" state due to the crystallographic inequivalence of the two $Fe^{3+}$ sites despite the same numbers of antiparallel spins. Different local coordination environments with distinct orbital hybridization with surrounding $O^{2-}$ could provide the finite orbital angular momentum of $Fe^{3+}$ or a different expectation value of the spin moment leading to different sizes of antiparallel magnetic moments even though the resulting slight difference between the two $Fe^{3+}$ magnetic moments may not easily be detectable by neutron scattering. Note that such orbital hybridization can correlate to the reported mixing of the higher-energy multiplet states [7] as observed in cobaltites [14,15]. This orbital mixing is predicted to stabilize the $A$ phase as discussed above [7] and can create a magnetic anisotropy even though the high-spin state of $Fe^{3+}$ generally exhibits a negligible magnetic anisotropy due to the absence of orbital angular momentum.

X-ray magnetic circular dichroism (XMCD) is a powerful and beneficial spectroscopic technique for the study of magnetism. It is possible to measure element- and site-specific magnetic hysteresis curves [16] and to extract orbital angular momentum from the overall magnetic moment of an atom by applying the sum rule [17]. Even if the finite orbital angular momentum is too small to be detected, distinct local coordination environments between the two sites can be reflected in the spectrum, enabling us to independently quantify the two almost equivalent magnetic sites.

Here, we report our investigation of the origin of $M$ in epitaxial $CaFe_2O_4$ films by means of synchrotron-based x-ray techniques, resonant elastic x-ray scattering (REXS) and XMCD. These two techniques are complementary since REXS on the (001) reflection, which is forbidden due to the $2_1'$ symmetry along [001], probes the AFM behavior from the $B$ phase without any contamination from the $A$ phase because the $A$ phase does not break the $2_1'$ symmetry along [001], while XMCD probes $M$ that could originate from the potentially "ferrimagnetic" $A$ phase. Our results indicate that the $B$ phase remains down to the lowest temperature with a developing population of the $A$ phase upon cooling, which resembles the behavior of single crystals. XMCD spectra show three contributions in $M$; two of them representing antiparallel $Fe^{3+}$ moments from the Fe1 and Fe2 sites, as expected for the $A$ phase, with the third one originating from $Fe^{2+}$, likely arising from defects, e.g., oxygen vacancies. Even though the orbital angular momentum that differentiates the size of the magnetic moments between the two sites is too small to be detected within the experimental precision, the measured magnetic hysteresis curves evidently attest to the ferrimagnetic nature of the $A$ phase.

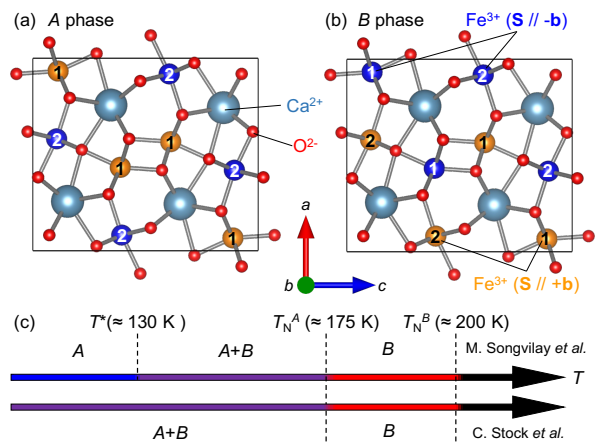

Fig. 1 Magnetic structures of $CaFe_2O_4$ in (a) the $A$ phase and (b) the $B$ phase, where orange and blue spheres represent $Fe^{3+}$ having a magnetic moment pointing to +$\mathbf{b}$ and –$\mathbf{b}$, respectively, and (c) its phase sequence for powder samples (up) [9] and for single-crystal samples (down) [8]. A number written on the spheres denotes the Fe sites, either Fe1 or Fe2. (a) and (b) were drawn by VESTA [28].

Epitaxial films of $CaFe_2O_4$ with a thickness of ~100 nm were grown on a $TiO_2$ (110) substrate by the pulsed laser deposition method. The films contain complex needle-like crystallines that organize forming domains because of the strain release processes and the epitaxial relation with the substrate. Details are described in Ref. [10]. Essentially for our study, there is a significant fraction of [001]-oriented domains along the surface normal, as displayed in Fig. 2(a). REXS experiments were performed at the RESOXS end-station [18] of the X11MA beamline [19] in the Swiss Light Source (Switzerland) and at the BL17SU [20] in the SPring-8 (Japan), and XMCD experiments were performed at the X07MA beamline [21] of the Swiss Light Source. The incident photon energy was set around the Fe $L_{2,3}$ edges for both types of experiments, whose setups are shown in Figs. 2(b) and 2(c). XMCD spectra were measured at several temperatures below $T_N{}^B$ in $H$ up to ±6 T. The grazing incidence of x rays (~30°) allows us to detect $M$ along [010], which is along the surface plane. We employed the total electron yield (TEY) and x-ray excited optical luminescence (XEOL) [22] detection modes, which provide surface- (~5-10 nm) and complete film depth sensitivity, respectively. To suppress charging affecting the TEY signals, 1.9 nm of Pt were deposited on the film surface. Representative spectra comparing TEY and XEOL data show excellent agreement (Supplemental Fig. S1). Due to the relatively large sample thickness, apparent self-absorption effects [22] hinder quantitative analysis of the XEOL spectra and hence, we show here only data taken with the TEY method.

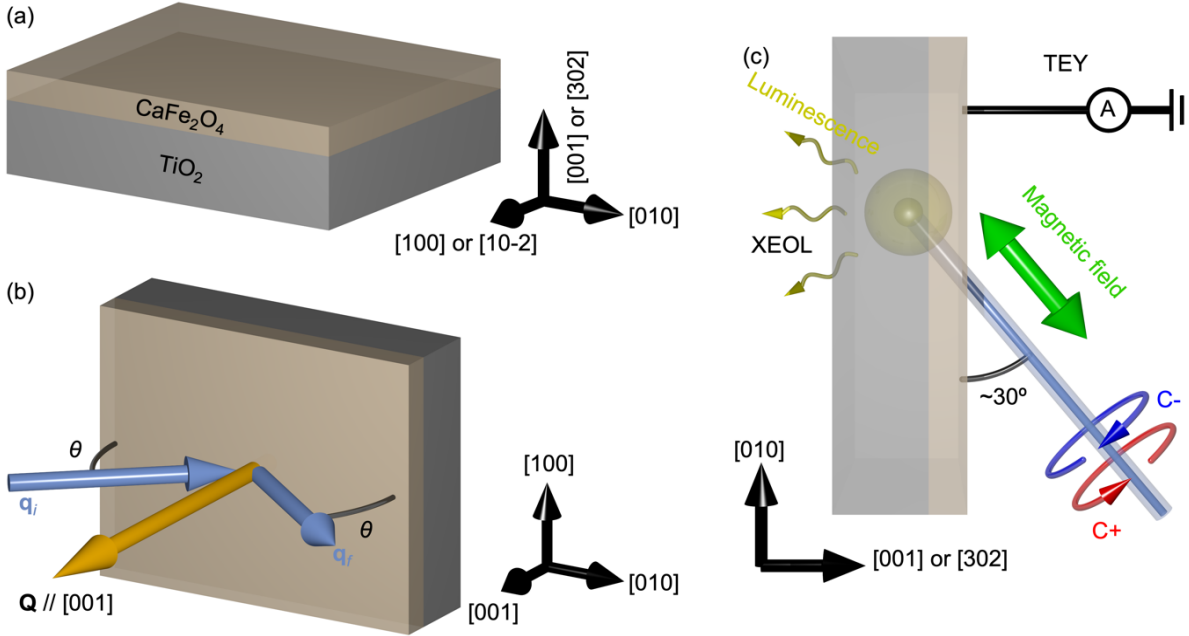

Fig. 2 (a) Schematic representation of the CaFe$_2$O$_4$ film orientation grown on a TiO$_2$ substrate, adapted from Ref. [10]. There are different crystallographic domains present, and the surface normal is either [001] or [302] oriented. Experimental setup for (b) REXS and (c) XMCD measurements. The (001) reflection appears around ~55° of $\theta$ at the Fe $L_3$ edge. For the XMCD measurements, we employed grazing incidence of x rays around 30°, enabling us to examine magnetization along [010].

Figure 3(a) shows (00$L$) REXS profiles around $L = 1$ at various temperatures. A small intensity observed even above $T_N^B$ is due to resonantly-allowed scattering from aspheric electron distribution ascribed to Fe $3d$ orbitals (see Supplemental Material for details). The presence of aspheric electron distribution is also confirmed from the observed quadrupole splitting in Mössbauer spectra [10]. The increase in the (001) reflection intensities for decreasing temperatures below $T_N^B$ [see Fig. 3(b)] reflects the growth of the $B$ phase. Note that the $A$ phase does not contribute to the (001) reflection because of the $2_1'$ symmetry along [001]. The distinct origins of the scattering below or above $T_N^B$ are also reflected in the differences in the energy spectra, displayed in Fig. 3(c). The presence of the $B$ phase down to the lowest achieved temperatures resembles the behavior of single-crystal samples, which show the coexistence of the $A$ and $B$ phases below $T_N^A$ [see Fig. 1(c)] [8].

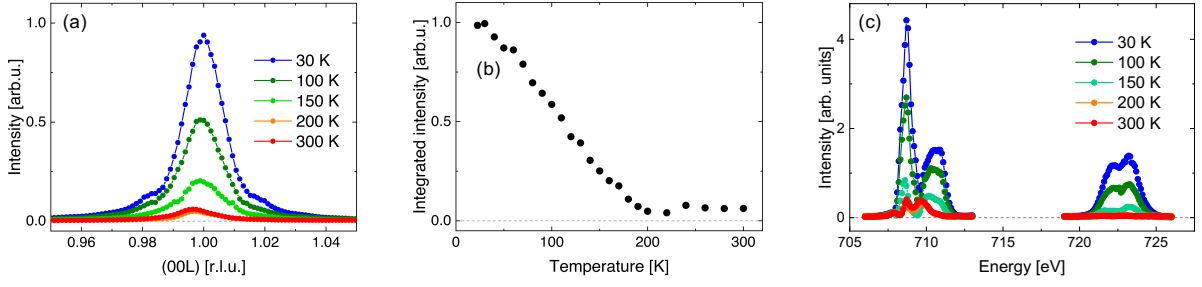

Fig. 3 (a) REXS profiles of the (001) reflection taken at various temperatures. (b) Temperature dependence of (001) integrated intensities, and (c) photon-energy dependence while maintaining the diffraction condition for the (001) reflection at various temperatures. (a) and (b) were taken at the photon energy of 710.8 eV.

To visualize how the $B$ phase develops upon cooling, we created two-dimensional maps of the (001) intensities at several temperatures across $T_N{}^B$ with an x-ray beam size of ~15 μm × 30 μm scanning with a step size of ~7 μm and ~15 μm for horizontal and vertical directions, respectively, as displayed in Figs. 4(a)-4(d). Because of the various crystallographic domains in the $CaFe_2O_4$ film [10], we can only sample the crystallographic domain fraction that has the [001] axis out-of-plane. Therefore, the intensities are spatially inhomogeneous even at 200 K in the absence of magnetic order, as seen in Fig. 4(d). The irregular shape of the intensity distribution is consistent with the previously reported structural domain morphology [10]. To better examine the developing magnetic contributions, we normalized the intensities taken at each temperature to those taken at 200 K. Figures 4(e)-4(g), obtained from the maps in Figs. 4(a)-4(c), establish that the magnetic contribution develops everywhere over the scanned region but is not uniform. The nonuniform magnetic profiles can be due to (i) the different amplitudes of the magnetic contributions from in-plane crystallographic domains [10] and/or (ii) the spatially inhomogeneous evolution of the $B$ phase. The latter implies the evolution of the $A$ phase, which does not contribute to the (001) intensities, and is consistent with the XMCD data shown later. Note that (ii) is not caused by inhomogeneously distributed crystallographic domains with a different out-of-plane orientation because of the normalization by the 200 K data.

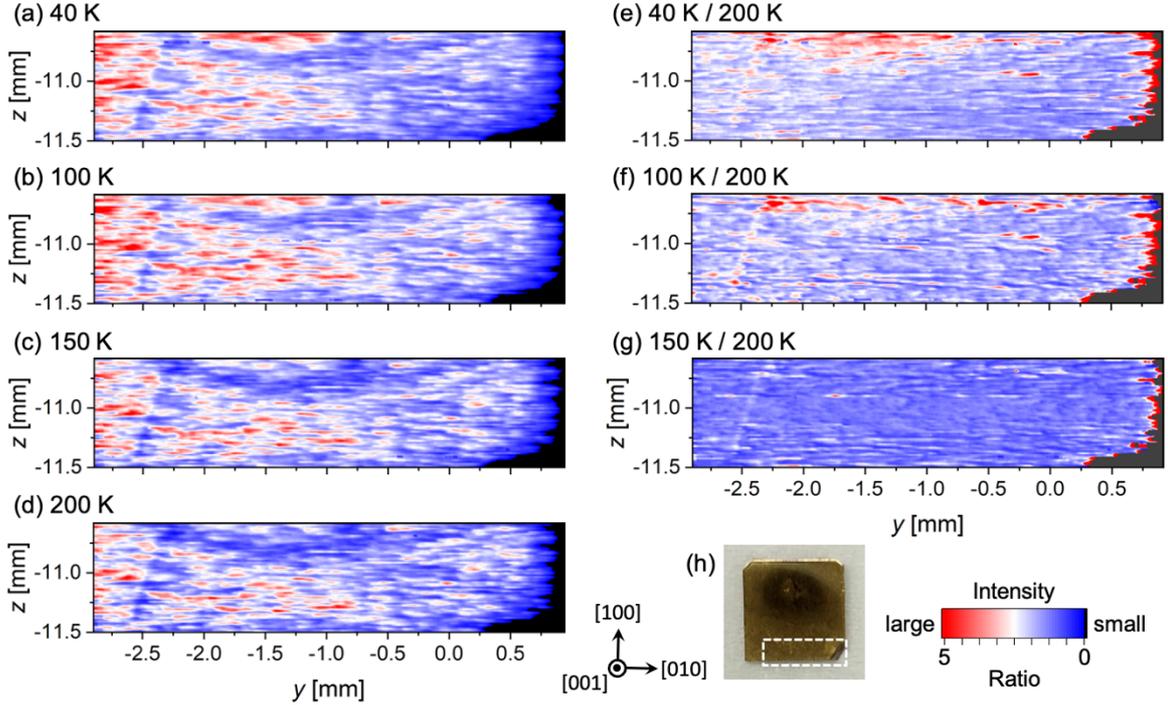

Fig. 4 Two-dimensional maps of (001) intensities taken at (a) 40 K, (b) 100 K, (c) 150 K, and (d) 200 K, with the photon energy of 710.25 eV. Intensities are normalized by those of the 200 K data at (e) 40 K, (f) 100 K, and (g) 150 K. These maps were taken in the region surrounded by a white box in (h). Black areas in (a)-(g) lie outside of the sample.

The appearance of the "ferrimagnetic" $A$ phase can be investigated by XMCD. Figure 5 displays XAS and XMCD spectra taken at $T$ = 30 K and 150 K in $H$ = 6 T after zero-field cooling (ZFC) from room temperature. The overall XAS with two well-resolved peaks ($e_g$ and $t_{2g}$) at the Fe $L_3$ edge is typical of a predominant $Fe^{3+}$ character. As the two different $Fe^{3+}$ sites are AFM coupled in the $A$ phase and are expected to have slightly different spectral shapes due to the distinct local coordination environments, the two peaks with opposite signs ($\beta$ and $\gamma$) in the XMCD can be assigned to these two sites. While the amplitude of $\gamma$ is similar between 30 K and 150 K, that of $\beta$ is different at the two temperatures and is comparable with the $\gamma$ peak at 30 K, further supporting their origins from distinct Fe-sites. Note that we found a history dependence on XMCD spectra for the two peaks, consistent with the fact that the $A$ phase is in competition with the $B$ phase and thus the $A$-phase population can differ by the $T$ and $H$ history conditions. Shown in Fig. 6 are the XMCD spectra taken at the same measurement condition, at 150 K in 6 T, but after approaching with different $T$ and $H$ evolution; after cycling $H$ (±6 T) at 2 K for the black curve, compared with no prior magnetic cycle for the green curve. Both the peaks are weaker in the green curve, indicating a smaller $M$ and less population of the $A$ phase with respect to the $B$ phase. The history dependence is consistent with the inhomogeneous development of the $B$ phase implied by the REXS data.

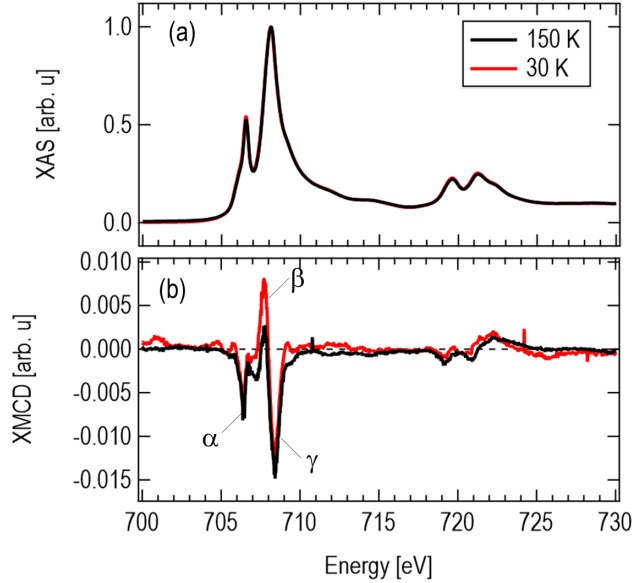

Fig. 5 (a) XAS and (b) XMCD taken at 30 K (red) and 150 K (black). XAS is obtained as the sum of two data with opposite helicity of circular polarization (C+/C-) and is normalized to its highest peak, whereas XMCD is obtained as the difference between C+ and C-.

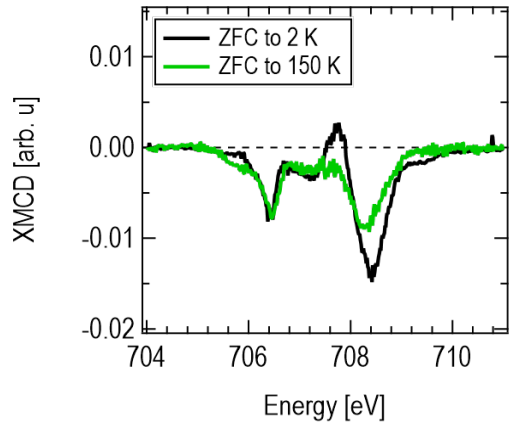

Fig. 6 History dependence of XMCD spectra taken at 150 K after ZFC directly from room temperature (green) compared with the XMCD at 150 K after ZFC to 2 K, magnetic cycling (±6 T), and warming in +6 T (black). The black data is the same as Fig. 5(b) and Fig. 7(b).

For the assignment of the XMCD spectral features, multiplet simulations of the Fe sites were performed using the MultiX code [23]. Crystal field multiplet parameters of a typical Fe $O_h$ site were used [24], but the local structure of the Fe1 and Fe2 sites was obtained from the bulk structure [25] in order to account for the differing octahedral distortions of the two crystallographic sites. Figure 7 shows the results of the multiplet simulations for the XMCD data shown in Fig. 5. The calculated site-selective XMCD spectra are shown in Figs. 7(c) and 7(d). Based on these results, we find that the Fe2 and Fe1 sites display a relative energy shift of ~0.45 eV accounting for the well-resolved β and γ peaks, respectively. On the other hand, the feature labeled as α is not reproduced well by the two $Fe^{3+}$ sites. Such a lower-energy peak compared to that of $Fe^{3+}$ is often assigned to $Fe^{2+}$. However, multiple possible contributions from anisotropic $Fe^{2+}$ in differently oriented domains hindered us to reproduce the feature uniquely with multiplet simulations. Nevertheless, the independent behavior of the peak α from the $Fe^{3+}$ features is clearly visible in the experiment, as shown in Fig. 6. Thus, it is reasonable to attribute the peak α to $Fe^{2+}$. Our observation of an XMCD signal from $Fe^{2+}$ suggests a high-spin state, unlike the proposed low-spin state ($S = 0$) for the $Fe^{2+}$ from magnetic susceptibility measurements of bulk samples [11]. $Fe^{2+}$ can originate from oxygen vacancies as discussed previously [11] and can form a ferrimagnetic cluster with surrounding $Fe^{3+}$. Note that the population of $Fe^{2+}$ must be very small as the XAS is dominated by the typical $Fe^{3+}$ spectra, which is also consistent with Mössbauer spectroscopy reports of a single order parameter of the $Fe^{3+}$ sites below $T_N{}^B$ [10,26] and no observable $Fe^{2+}$ (detectable typically within few percent).

The peak β reflecting the $Fe^{3+}$ moments of the Fe2 site is larger at 30 K than at 150 K while the peak γ reflecting the $Fe^{3+}$ moments of the Fe1 site is almost the same between the two temperatures, indicating different temperature evolution of the Fe1 and Fe2 sites in the $A$ phase. At +6 T and 150 K, we find the antiparallel orientation of the Fe1 and Fe2 sites with a clear imbalance of the site-specific magnetization with $M_{Fe1} > M_{Fe2}$. At +6 T and 30 K, $M_{Fe2}$ becomes larger and both $Fe^{3+}$ sites contribute approximately equally to the XMCD. This can be understood by the growth of the $A$ phase when lowering the temperature, as reported in single-crystal samples; for higher $T$ the Fe2-site moments (β) are significantly smaller than the Fe1-site moments (γ) in thin domains of the $A$ phase because of the imbalanced population, whereas for lower $T$ the population gets more balanced [see the sketch in Fig. 7(e)]. In other words, it suggests that the intrachain exchange interaction between two Fe1 is likely more FM than that between two Fe2. The favorable FM interaction between two Fe1 might correlate with more distortion in the octahedral coordination than Fe2, which can mix the higher-energy multiplet states into the ground state and stabilize the $A$ phase with intrachain FM coupling between the same sites of Fe. The view in the sketch explains larger $M$ at 150 K than 30 K [10] as two net $M$ from the sublattices represented by the peaks β and γ are antiparallel to each other and β develops more and gets comparable with γ at lower temperatures. Since the local FM spin configuration in the $A$ phase is the same as that in the antiphase domain boundary of the AFM $B$ phase, the slightly different onset temperature of $M$ (~185 K) compared to $T_N{}^A$ might imply the short-range order of the $A$ phase, which is likely present as the antiphase domain boundary of the $B$ phase.

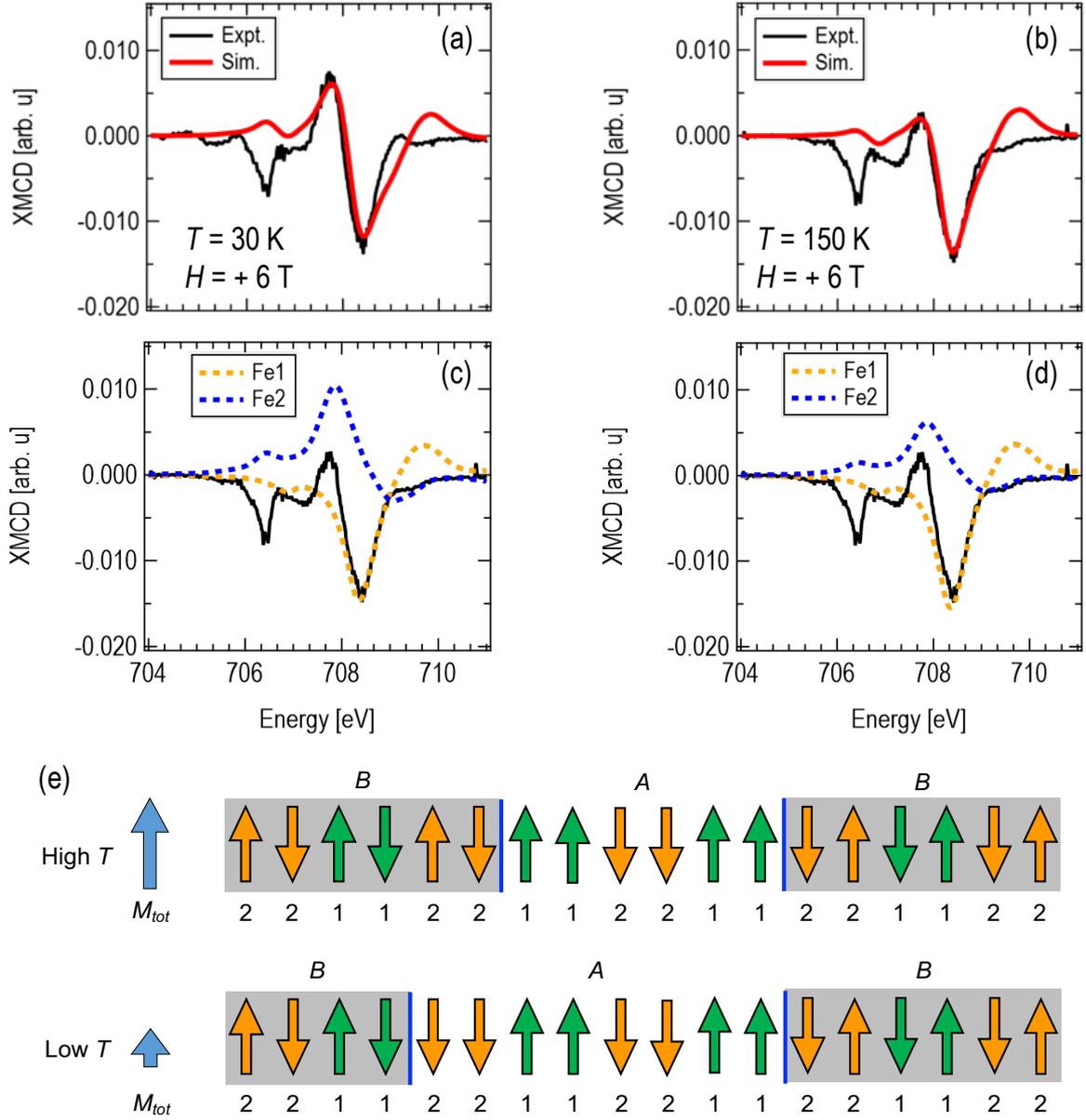

Fig. 7 XMCD spectra taken at (a) 30 K and 6 T, and (b) 150 K and 6 T, which are decomposed into three contributions, as found in (c) and (d). Broken curves correspond to calculated respective contributions to the spectra while a red curve corresponds to the sum of the contributions. (e) A sketch representing the growth of the A phase and resultant decrease in total magnetization.

We aimed to obtain direct evidence that the A phase is ferrimagnetic. The reversal of β and γ XMCD peaks should take place by sweeping H. The same fitting procedure as done for the data in Fig. 7 enables us to extract site-specific magnetic hysteresis curves. Figures 8(a) and 8(b) display XMCD spectra taken at 150 K and 30 K, respectively, in various H. Extracted magnetic hysteresis curves of the two Fe sites are shown in Figs. 8(c) and 8(d). It is observed that the XMCD peaks reverse their signs by sweeping H at 150 K, directly evidencing that these two AFM-coupled $Fe^{3+}$ sites are responsible for $M$, namely the $A$ phase of $CaFe_2O_4$ is ferrimagnetic.

Another feature is that $M$ from the two $Fe^{3+}$ sites does not reverse at 30 K even when going from +6 T to –6 T, in contrast to the data taken at 150 K. This observation is counterintuitive as the high-spin configuration of $Fe^{3+}$ commonly exhibits a small anisotropy leading to small coercive fields due to the equal filling of the orbitals. However, the tiny overall magnitude of $M$ from the two imperfectly compensated sublattice magnetizations leads to a strongly reduced force on $M$ from $H$, naturally explaining an enlarged coercive field. In addition, the significant distortion of the $FeO_6$ octahedron that could mix higher-energy multiplet states into the ground state through charge transfer (or orbital hybridization) with the surrounding $O^{2-}$ [7] may generate magnetic anisotropy that results in an enlarged coercive field.

The minor steps observed at $H = 0$ T in the 30 K $Fe^{3+}$ data of Figs. 8(c) and 8(d) can be due to (i) the flipping of $Fe^{2+}$ XMCD signals that spectrally overlap on the peaks β and γ and/or (ii) the flipping of $Fe^{3+}$ moments coupling to $Fe^{2+}$. Both possibilities imply a small coercive field of $Fe^{2+}$ moments. Therefore, the uncoupled spins from the AFM phase, i.e., from the main $Fe^{3+}$ spins, reported previously [13] should be ascribed to defective $Fe^{2+}$ that can easily respond to $H$.

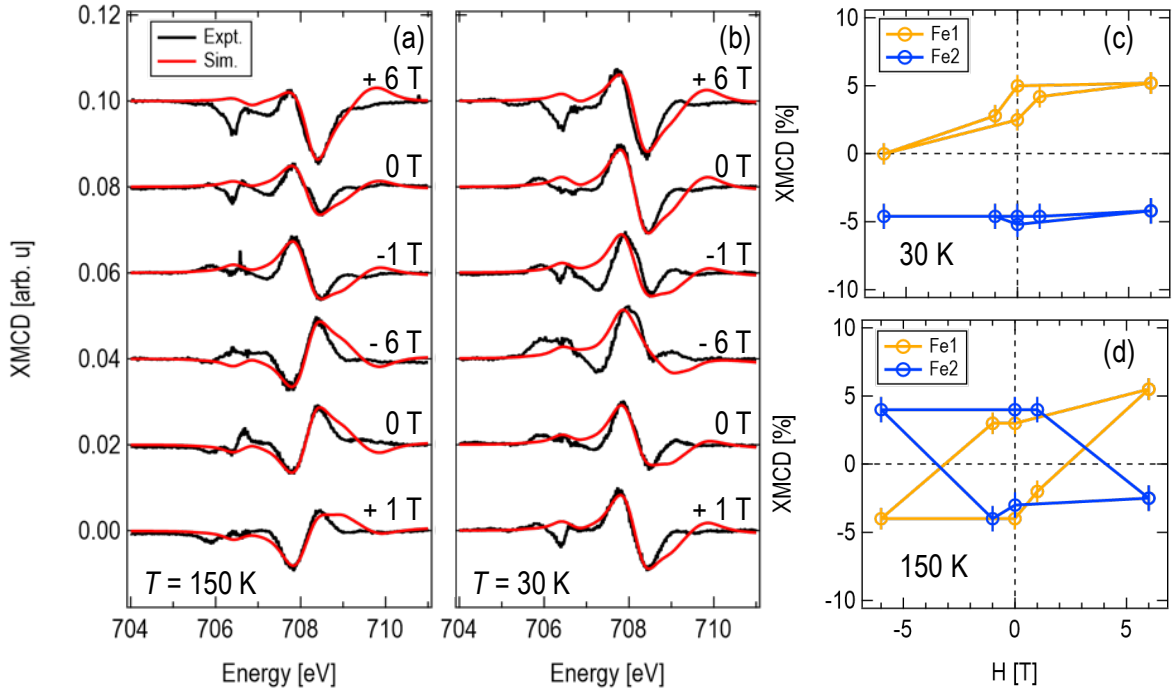

Fig. 8 Magnetic-field dependence of the XMCD signals at (a) 150 K and (b) 30 K. Hysteresis curve for $Fe^{3+}$ (1) and $Fe^{3+}$ (2) at (c) 30 K and (d) 150 K obtained from fits with calculated spectra to (a) and (b).

In conclusion, we have investigated the origin of the tiny remanent magnetization reported in nominally antiferromagnetic $CaFe_2O_4$ by synchrotron-based spectroscopic techniques. Although the numbers of antiparallel spins, up and down both hosted by $Fe^{3+}$ with octahedral coordination, are exactly the same in the $A$ phase, we clarified that the $A$ phase is ferrimagnetic due to the difference in the local coordination environments of the two sites by measuring site-specific magnetic hysteresis loops. The magnetization magnitude is comparably small with an independent defect contribution ($Fe^{2+}$) having a different coercive field. Such a ferrimagnetic state may be useful for future spintronics as (i) the coercive field is much larger than normal ferromagnets or ferrimagnets due to small overall magnetization, as evident in our study on $Fe^{3+}$, which usually exhibits small coercive fields, demonstrating the robustness of its magnetic state under an external magnetic field, (ii) there are two switchable ferroic states ($\pm M$), and (iii) dominating antiferromagnetic interactions, allowing magnetic resonance frequencies in the sub-THz range as antiferromagnets. Our discovery may open up a new direction of material design for future devices.

**Data availability**

Experimental data are accessible from the PSI Public Data Repository [27].

**Acknowledgements**

The resonant x-ray diffraction experiments were performed at the X11MA beamline in the Swiss Light Source under proposal No. 20191307 and at the BL17SU in the SPring-8 under proposal No. 20200012. The x-ray magnetic circular dichroism measurements were performed at the XTreme beamline in the Swiss Light Source during in-house access. H.U. acknowledges the National Centers of Competence in Research in Molecular Ultrafast Science and Technology (NCCR MUST-No. 51NF40-183615) from the Swiss National Science Foundation and from the European Union's Horizon 2020 research and innovation programme under the Marie Skłodowska-Curie Grant Agreement No. 801459 – FP-RESOMUS. E.S. is supported by the NCCR Materials' Revolution: Computational Design and Discovery of Novel Materials (NCCR MARVEL No. 182892) from Swiss National Foundation and the European Union's Horizon 2020 research and innovation programme under the Marie Skłodowska-Curie Grant Agreement No. 884104 (PSI-FELLOW-III-3i). N. O. H. acknowledges financial support of the Swiss National Science Foundation, No. 200021_169017. M.B. is supported by the Swiss National Science Foundation through project Nos. 200021-196964 and 200021_169698, respectively. Financial support by the Groningen Cognitive Systems and Materials Center (CogniGron) and the Ubbo Emmius Funds of the University of Groningen is also gratefully acknowledged.

**Additional information**

**Competing interests**: The authors declare no competing interests.